\documentclass[journal=jacsat,manuscript=article]{achemso}

\usepackage[utf8]{inputenc}
\usepackage{graphicx}
\usepackage{wrapfig}
\usepackage[colorlinks,allcolors=black,citecolor=blue,urlcolor=blue]{hyperref}
\usepackage[mathlines]{lineno}
\usepackage{ae}
\usepackage{amsmath}
\usepackage{booktabs}
\usepackage{silence}
\usepackage[dvipsnames]{xcolor}
\usepackage[symbol]{footmisc}

\usepackage[version=3]{mhchem} 

\usepackage{amsmath,amssymb}

\DeclareFontFamily{U}{rcjhbltx}{}
\DeclareFontShape{U}{rcjhbltx}{m}{n}{<->rcjhbltx}{}
\DeclareSymbolFont{hebrewletters}{U}{rcjhbltx}{m}{n}
\DeclareMathSymbol{\shin}{\mathord}{hebrewletters}{152}

\author{H. Christian Schewe}
\affiliation{Fritz-Haber-Institut der Max-Planck-Gesellschaft, Faradayweg 4-6, 14195 Berlin, Germany}
\alsoaffiliation{Institute of Organic Chemistry and Biochemistry, Czech Academy of Sciences, Flemingovo nám. 2, 16610 Prague 6, Czech Republic}
\altaffiliation{These authors contributed equally}

\author{Bruno Credidio}
\affiliation{Institute for Chemical Sciences and Engineering (ISIC), Ecole Polytechnique F\'ed\'erale de Lausanne (EPFL), 1015 Lausanne, Switzerland}
\altaffiliation{These authors contributed equally}

\author{Aaron M. Ghrist}
\affiliation{Fritz-Haber-Institut der Max-Planck-Gesellschaft, Faradayweg 4-6, 14195 Berlin, Germany}
\alsoaffiliation{Department of Chemistry, University of Southern California, Los Angeles, CA 90089-0482, USA}
\altaffiliation{These authors contributed equally}

\author{Sebastian Malerz}
\affiliation{Fritz-Haber-Institut der Max-Planck-Gesellschaft, Faradayweg 4-6, 14195 Berlin, Germany}

\author{Christian Ozga}
\affiliation{Institut für Physik und CINSaT, Universität Kassel, Heinrich-Plett-Straße 40, 34132 Kassel, Germany}

\author{André Knie}
\affiliation{Institut für Physik und CINSaT, Universität Kassel, Heinrich-Plett-Straße 40, 34132 Kassel, Germany}

\author{Henrik Haak}
\affiliation{Fritz-Haber-Institut der Max-Planck-Gesellschaft, Faradayweg 4-6, 14195 Berlin, Germany}

\author{Gerard Meijer}
\affiliation{Fritz-Haber-Institut der Max-Planck-Gesellschaft, Faradayweg 4-6, 14195 Berlin, Germany}

\author{Bernd Winter}
\affiliation{Fritz-Haber-Institut der Max-Planck-Gesellschaft, Faradayweg 4-6, 14195 Berlin, Germany}

\author{Andreas Osterwalder}
\email{andreas.osterwalder@epfl.ch}
\affiliation{Institute for Chemical Sciences and Engineering (ISIC), Ecole Polytechnique F\'ed\'erale de Lausanne (EPFL), 1015 Lausanne, Switzerland}

\title{Imaging of chemical kinetics at the water–water interface in a free-flowing liquid flat-jet}

\begin{document}

\begin{abstract}
We present chemical kinetics measurements of the luminol oxydation chemiluminescence reaction at the interface between two aqueous solutions, using liquid jet technology.
Free-flowing iquid microjets are a relatively recent development that has found its way into a growing number of applications in spectroscopy and dynamics.
A variant thereof, called flat-jet, is obtained when two cylindrical jets of a liquid are crossed, leading to a chain of planar leaf-shaped structures of the flowing liquid.
We here show that in the first leaf of this chain the fluids do not exhibit turbulent mixing, providing a clean interface between the liquids from the impinging jets.
We also show, using the example of the luminol chemiluminescence reaction, how this setup can be used to obtain kinetics information from friction-less flow and by circumventing the requirement for rapid mixing but by intentionally suppressing all turbulent mixing and instead relying on diffusion.

\end{abstract}

\section{Introduction}
\begin{figure}[H]
    \centering
    \includegraphics[width=\linewidth]{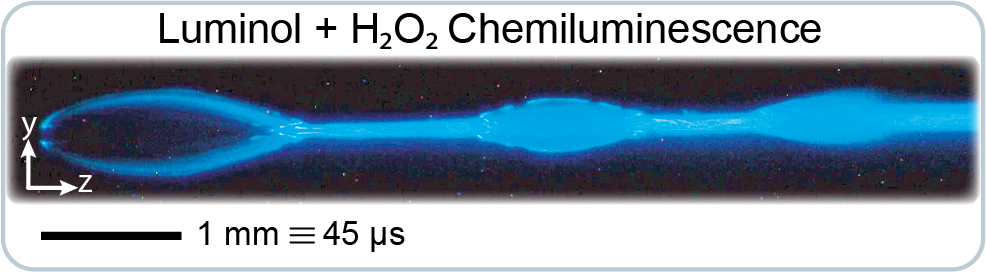}
    \caption{Photograph of a water flat-jet glowing blue as a result of chemiluminescence.}
    \label{fig:glow}
\end{figure}

Fast-flowing liquid microjets are a powerful tool for the preparation of volatile liquids, including water, even in high-vacuum environments.\cite{faubelPhotoelectronSpectroscopyLiquid2000,faubelPhotoelectronSpectroscopyLiquid1997,faubelUltrafastSoftXray2012,winterPhotoemissionLiquidAqueous2006,faubelMolecularBeamStudy1988}
They recently gained much interest in particular for in-vacuum applications where the jet travels freely for some millimeters before decaying into droplets and freezing. 
One of the prime factors that make them interesting tools is the free flow through air or vacuum which permits unobstructed optical access to the liquid and thus enables a wide range of spectroscopic detection that are incompatible with many solid container materials.
Recent applications include X-ray photoelectron spectroscopy,\cite{aliElectronicStructureAqueousphase2019,buttersackValenceCoreLevelXray2019,franssonXrayElectronSpectroscopy2016,jungwirthIonsAqueousInterfaces2008,lewisCO2CaptureAmineBased2011,winterLiquidMicrojetPhotoelectron2009,karashimaUltrafastInternalConversion2019,suzukiTimeresolvedPhotoelectronSpectroscopy2012,suzukiUltrafastPhotoelectronSpectroscopy2019,faubelPhotoelectronSpectroscopyLiquid2000}evaporation dynamics,\cite{ryazanovQuantumstateresolvedStudiesAqueous2019,faustGasMicrojetReactive2016,hahnSuperMaxwellianHeliumEvaporation2016,lancasterProbingGasLiquid2015,sobyraLiquidMicrojetMeasurements2017,murdachaewDeprotonationFormicAcid2016,faustMicrojetsCoatedWheels2016a}
attosecond-pulse generation,\cite{jordanPhotoelectronSpectrometerLiquid2018,yinFewcycleHighharmonicGeneration2020}
and liquid-gas scattering.\cite{lancasterInertGasScattering2013,lancasterInertGasScattering2013,artigliaSurfacestabilizedOzonideTriggers2017}
 The most common implementation is a single cylindrical jet, obtained by forcing the liquid at a pressure of a few bars through a 10-50 $\mu$m-diameter nozzle, which results in a laminar jet with a flow velocity of tens of m/s.
 
Many experiments demand a planar surface in order to avoid unwanted averaging over effects resulting from the angle-dependent surface normal.\cite{thurmerPhotoelectronAngularDistributions2013,lancasterInertGasScattering2013}
Different arrangements exist to produce laminar-flow planar surfaces,\cite{galinisMicrometerthicknessLiquidSheet2017,koralekGenerationCharacterizationUltrathin2018}
among which a widely used setup is the crossing and impinging of two cylindrical jets.\cite{ekimovaLiquidFlatjetSystem2015} 
For large enough Reynolds number this produces a chain of few-micron thin leaf-shaped sheets, each bound by a relatively thick fluid rim and stabilized by an interplay of surface tension and fluid inertia. 
Consecutive sheet planes are perpendicular to each other. 
The stability and geometry of this structure are governed by solution properties such as surface tension and viscosity, and by controlled parameters like flow-rate and jet-diameter.\cite{chenHIGHFIDELITYSIMULATIONSIMPINGING2013,ekimovaLiquidFlatjetSystem2015}

An important question about such objects is wether the first leaf of the chain contains a turbulent mixture of the fluids from the two jets, or if these flow alongside each other, yet we are not aware of any experimental study addressing this question. 
Indeed, the second option implies that a well-defined liquid--liquid interface is generated which is of great interest in the case of two different solvents, but also for identical solvents of different composition. 
Previous studies have shown that in microfluidics devices it is possible to prepare well-defined interfaces between miscible and immiscible fluids by keeping the liquid flow laminar, and these devices have been used in studies targeting structural and dynamical aspects of interfaces.\cite{atenciaControlledMicrofluidicInterfaces2005,ismagilovExperimentalTheoreticalScaling2000a}
In contrast to free-flowing flat jets, however, microfluidics inherently require a container material which, on the one hand, imposes limitations on the systems in terms of flow dynamics, since friction on the walls leads to modified flow patterns, and on the available spectroscopic tools on the other hand, since the container material itself absorbs electromagnetic radiation in ranges that may be critical for the system under investigation.

We here demonstrate that impinging, but free-flowing, jets do produce a leaf structure where the fluids flow alnogside each other in the first leaf and thus represent a tool to gain access to the liquid-liquid interface of miscible fluids. 
This finding demonstrates an important aspect of free-flowing microjets which makes them a powerful borderless alternative to microfluidics and opens the door towards the investigation of chemical reactions and in-vacuum studies at liquid-liquid interfaces, with the option of using extreme ultraviolett and X-ray radiation that are only limited by the absorption spectrum of the solvent itself.
Species in either solution diffuse across the interface while flowing downstream, thereby creating a steady-state system with an increasingly overlapping region where chemical reactions can take place. 
Based on this we here also demonstrate a technique for chemical kinetics studies under completely controlled conditions which explicitly avoids the necessity for rapid mixing and benefits from the free-flowing jets that are not perturbed by friction on container-walls.\cite{songMillisecondKineticsMicrofluidic2003}
By combining two jets with different reactants and applying a suitable spectroscopic detection scheme, the flow axis of the jet represents the time axis, permitting to directly see and image the progress of the reaction.
Typical leaf surface dimensions are 1.5 mm x 0.5 mm, and with flow velocities of a few 10 m/s this corresponds to a flow time through the first leaf of around 50--100 $\mu$s, thus covering a time scale for chemical kinetics that currently is difficult to access by other methods.

The present experiment detects photons from a chemiluminescence (CL) reaction, allowing at the same time the proof for the controlled formation of the liquid--liquid interface, and the imaging of the chemical reaction kinetics.
Our sample reaction is the oxidation of 5-amino-2,3-dihydro-1,4-phthalazinedione, known as luminol, which is oxidized by \ce{H2O2} in the presence of a transition metal ion.\cite{roseChemiluminescenceLuminolPresence2001,matsumotoDiffusionMicrochannelAnalyzed2013}
The reaction product is an electronically excited state of deprotonated 3-aminophthalic acid ($AP^*$) which relaxes to the electronic ground state by emitting a blue photon. \cite{merenyiEquilibriumReactionLuminol1984,merenyiLuminolChemiluminescenceChemistry1990,merenyiREACTIVITYSUPEROXIDEO21985}

Injecting the principal reactants through individual jets yields a liquid flat-jet that luminesces exclusively where the liquids mix. 
Figure \ref{fig:glow} is the resulting photo of the glowing flat-jet, where the cylindrical jets enter from the left and flow to the right. 
Such an image provides a direct visualisation of the dynamics in the flat-jet: from the intensity distribution of the emitted light we can immediately conclude that there is no turbulent mixing in the flat region of the first leaf and infer information on the kinetics. 
Complete mixing of the solutions leads to a uniform blue glow as it is observed in rims and onward from the second leaf. 
However the first leaf displays a gradual increase of the luminescence, indicative of diffusion across the well-defined interface, and defined by the relative rates of  diffusion, chemical reaction kinetics, and luminescence.
The luminol–\ce{H2O2} reaction has been found to take place on the sub-millisecond time scale, while our imaging setup allows for a time resolution of around 2 $\mu$s (see SI). 
This system is thus ideally suited for spatial mapping of solution mixing and the demonstration of the feasibility of reaction kinetics measurements.

\begin{figure}[H]
    \centering
    \includegraphics[width=0.8\linewidth]{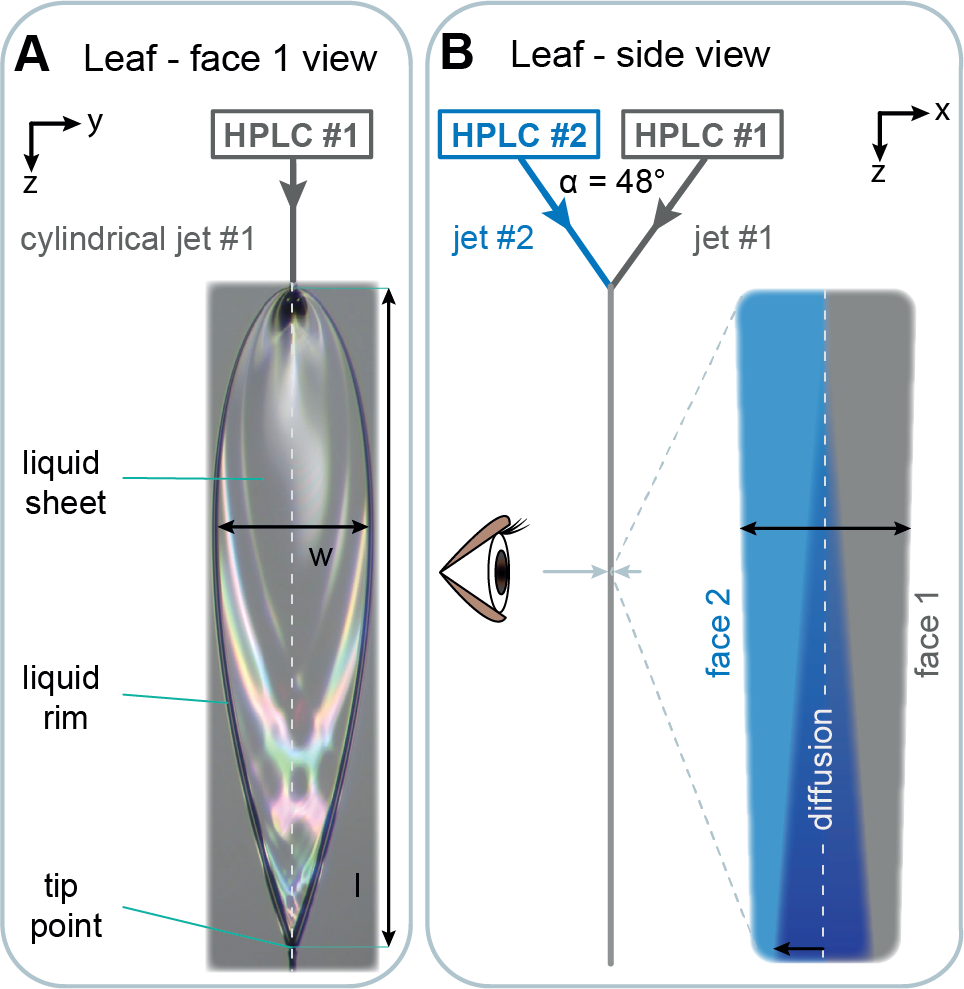}
    \caption{Experimental arrangement to generate the flat-jet. Two cylindrical liquid microjets, operated by separate HPLC pumps, are crossed at an angle of $\alpha$ = 48$^\circ$. One contains a luminol solution and Cu$^{2+}$ ions, the second an aqueous \ce{H2O2}. (A) View at the face of the first leaf. The plane spanned by the two cylindrical jets is vertical to the leaf plane. (B) Sketch showing the geometry used for CL detection and a cross section of the first leaf. A CCD camera faces the leaf surface and collects the CL.}
    \label{fig:setup}
\end{figure}
In the present experiment, described in detail in the SI, we generate the flat-jet by colliding two liquid jets in atmosphere, one containing an aqueous \ce{H2O2} solution, the other containing luminol and copper ions. 

A  photograph of a typical flat-jet, but with no chemical reaction, used in the present experiments is shown in Figure \ref{fig:setup}A. 
Rims and turn-over point to the second leaf are clearly visible. 
 Color structures on the leaf surface are optical interferences resulting from the sample illumination. 
 Figure \ref{fig:setup}B shows a cross section of the flat-jet in the plane of the two original jets, sketching the merged structure formed by the two flat-jets of solutions \#1 and \#2, joined by the interface. 
Here the two cylindrical jets are represented by the two arrows crossing at angle $\alpha$. 
 The downstream increase of the interfacial layer thickness, indicated as the dark blue area, results from diffusion of species across the boundary (see the SI for calculations of the resulting concentration distributions). 

\section{Analysis of Chemiluminescence Images}
\begin{figure}[H]
    \centering
    \includegraphics[width=0.5\linewidth]{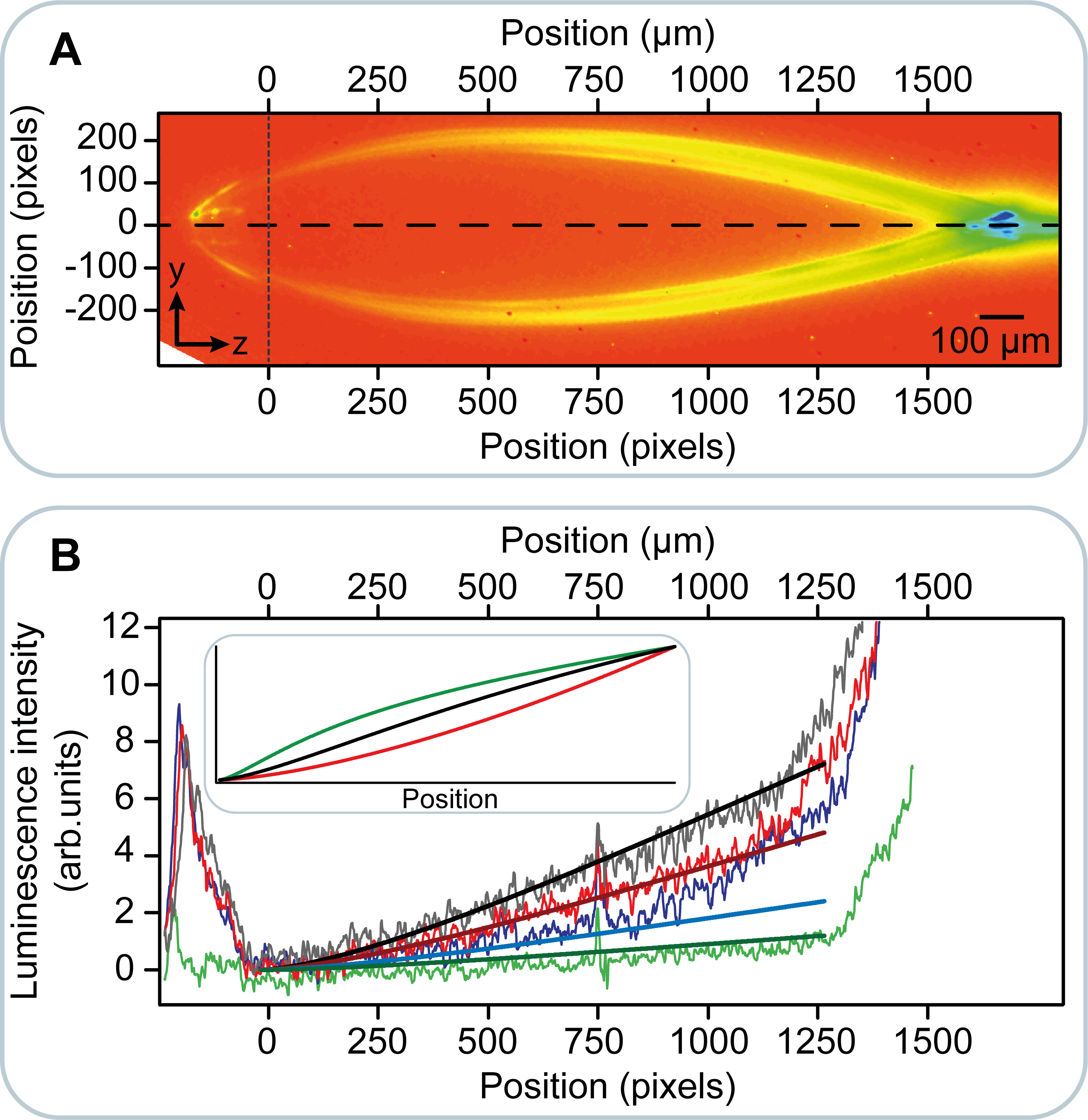}
    \caption{(A) False-color image of the measured CL intensity within the first leaf of the flat-jet (blue is high intensity; red is zero intensity). (B) CL signal intensity along the $y$-direction at $z$ = 0, indicated by the dashed horizontal line in A, for luminol concentrations of 6 g/L (black line), 4 g/L (red line), 2 g/L (blue line) and 1 g/L (green line), respectively. Thick solid lines are numerical results. Inset: calculated results for cases with extreme (red and green) ratios and the result found here (black) between reaction and diffusion rates.}
    \label{fig:results}
\end{figure}
Figure \ref{fig:results}A shows a false-color plot of the CL intensity in the $y-z$ plane, from the first leaf in Figure \ref{fig:glow}, with the camera oriented along the $x$-axis (see Figure \ref{fig:setup}B). 
Two 50 $\mu$m jets enter from the left, in the $x-z$ plane.
The first leaf in this arrangement has dimensions of l = 1.5 mm w = 0.3 mm, respectively, and a flow speed of 23 m/s.
They are not visible because no luminescence is emitted from the individual reactant jets which impinge at around $y$ =-100 $\pm$100 pixels and $z$ =0$\pm$50 pixels. 
Calibration of the pixel scale and conversion to the $\mu$m scale at the top is obtained using the known cylindrical-jet diameter in Figure \ref{fig:setup}A. 
The flat-jet origin is defined as the point where a laminar flat-jet forms and CL starts to be observable. 
CL intensity at negative z values and in the rims mirrors the turbulences in those regions.
In contrast, in the flat region only a slight intensity increase, starting at zero luminescence, is observed along the flow direction. 

The specific form of this increase is defined by the overall dynamics of the process and is used to extract information on the rates of the involved processes.
The CL intensity as a function of the downstream coordinate (along the $y$-axis at $z$ = 0; dashed line in Figure \ref{fig:results}A) is shown in Figure \ref{fig:results}B. 
Results are plotted for four different solutions with luminol concentrations of 6 g/L (grey), 4 g/L (red), 2 g/L (blue) and 1 g/L (green), respectively (all other reactants in this solution are scaled accordingly). 
These graphs, along with Figures \ref{fig:glow} and \ref{fig:results}A, reveal the following:
1. All leaves except the first one exhibit uniform CL intensity of similar magnitude in the flat parts and in the rims, indicating complete mixing of the two liquids. 
2. The CL intensity observed at $y<$0 first decreases to zero before gradually increasing. 
3. The CL intensity in the first leaf increases towards positive $y$, supporting the assumption of reagent diffusion across the interface. 
Point 2 is an important confirmation that the CL reaction, including the final luminescence step, proceeds fast on the time scale in comparison with flow speed, diffusion and reaction kinetics, and that at the origin a purely laminar flow is established (also see SI).

The overall kinetics are assessed by theoretically reproducing the measured CL profiles of Figure \ref{fig:results}B, using a simple kinetics model that includes the diffusion, the principal reaction steps (detailed in the SI), and luminescence. 
The photon flux $\Phi$ from CL results from the decay of $AP^*$:
\begin{equation}
\Phi=k_P [AP^*](t),\label{eq:phi}
\end{equation}
where $k_P$ is the rate coefficient for CL and $[AP^*](t)$ is the time-dependent product concentration. 
$AP^*$ is formed in a series of chemical reactions outlined in the SI.\cite{burdoMechanismCobaltCatalysis1975,gaikwadSelectiveStoppedflowDetermination1995,ojimaCANTFINDTHIS2002,merenyiLuminolChemiluminescenceChemistry1990,roseChemiluminescenceLuminolPresence2001}
The rate of this process can be written as a function of the limiting reactant concentrations
\begin{equation}
d\left[AP^*\right]/dt  =   k_R \left[\ce{H2O2}\right]\left[\ce{LH-}\right],\label{eq:rate}
\end{equation}	 		
where $k_R$ is a collective rate coefficient for the chemical reaction sequence that incorporates the various steps leading to $AP^*$.
The required high pH of the luminol solution indicates that it is present exclusively in the deprotonated form \ce{LH-}, conditions which have been found to be favorable for the reaction. \cite{menziGenerationSimpleCharacterization2020,osullivanStoppedFlowLuminol1995}

Applying the coordinate system introduced in Figures \ref{fig:setup} and \ref{fig:results}A, the relevant spatial dimensions thus are both the $y$-axis (which translates into a time axis, $t$) and the $x$-coordinate perpendicular to the leaf surface which is the principal direction of diffusion and hence mixing of the reactants. 
The concentration distribution $c(x,t)$ is derived from the initial concentrations, $c_0$, using Fick's law.\cite{steinfeldChemicalKineticsDynamics1998}
In the model of the two-solution flat-jet the initial concentration distributions, $c(x=0,t=0)$, are step-functions where the concentration of either reactant is $c_0$ on its original side and zero elsewhere. 
As time progresses, the diffusion-determined profile for each specie $i$ is described by an error-function $\Gamma$: 
\begin{equation}
c_i(x,t)=\frac{c_{0,i}}{2} \left\{1-\Gamma\left(\frac{x}{2\sqrt{D_{i}t}}\right)\right\},\label{eq:gamma}	
\end{equation}
where $D_{i}$ is the diffusion coefficient for specie $i$, and $t$ is time, defined by the flow velocity $v$ and position along the jet propagation as $t = y/v$. 

In a second step the concentration profiles from equation \ref{eq:gamma}, expressed as a function of $y$, are convoluted with each other and fed into equation \ref{eq:rate}, thus incorporating diffusion into the kinetic model. 
This yields the final profile for $[AP^*]$ along the $y$ axis, and equation \ref{eq:phi} is then used to calculate the CL intensity:
\begin{equation}	
\frac{\Delta\left[AP^*\right]}{\Delta y}=\frac{1}{v}\left[-k_P[AP^*]+k_R\shin(y)\right]\label{eq:APy}
\end{equation}	
Here, $\shin(y)$ is the integral along $y$ of the convoluted concentration profiles along $x$, and it thus accounts for the widening of the interface downstream in the jet (see SI). 
The terms in the brackets are the rate of decrease of $[AP^*]$ due to CL, in accordance with equation \ref{eq:phi}, and  of increase due to the chemical reaction, equation \ref{eq:rate}. 

The resulting expression is then numerically integrated to yield synthetic luminescence traces along the jet axis, and rate coefficients are fitted to reproduce the experimental results from Figure \ref{fig:results}B, as is shown by thick solid lines in the same graph.
In the current proof-of-principle study we are not able to extract absolute rate coefficients because some of the rates are correlated, and because in this case a calibration of the luminescence measurement was not attempted. 
Here we present a simplified analysis that reproduces the shape of the traces that depends on the relative magnitudes of the rates of diffusion, luminescence, and chemical kinetics.
In the SI we elaborate on a procedure that will enable a quantitative analysis for future experiments where a calibration will be included.

From the results in Figure \ref{fig:results}B we extract the following:
The interplay and relative magnitudes of diffusion, reaction kinetics, and fluorescence rates determine the overall magnitude of the CL signal but in particular also the profile of the intensity along the $y$ axis. 
The global shape of the curve is an asymmetric sigmoidal evolution, and depending on the relative magnitudes of the different rates we probe different parts of that curve, as shown in the inset of Figure \ref{fig:results}B. 
The curve starts with a positive curvature which is emphasized in the red curve where we artificially reduced the diffusion coefficients.
It converges, showing a negative curvature, on a threshold, as is clearly seen in the green curve where we artificially slowed down the luminescence.
The intermediate regime, shown in black, provides a near-linear evolution as is also observed experimentally.
Interestingly, the reaction rate coefficient itself merely leads to an overall scaling of the signal but without affecting the shape, thus underlining that for a qualitative understanding, calibrated measurements seem unimportant: even without it, the shape of the curve reveals principal aspects of the dominating kinetics.
The curvature also does not depend on the concentrations, and nearly identical results to those shown in Figure \ref{fig:results} are obtained by linearly scaling a single calculation with concentration.

\section{Conclusions}
A free-flowing liquid flat-jet has been produced from two different solutions to form a controlled liquid-liquid interface with uninhibited optical access from both sides.  
Using the CL reaction of luminol-oxydation by \ce{H2O2}, we created a marker for the temporal evolution of the mixing in the interfacial layer. 
By measuring the CL intensity distribution from the surface of the first leaf we demonstrate that the individual cylindrical jets merge into a laminar regime, and mixing between the two solutions within this leaf happens solely due to diffusion. 
We further show that the obtained CL image reveals reaction kinetics on the sub-ms timescale. 
We observe a linear dependence of CL on reactant concentrations, indicative of first order kinetics in the rate-limiting substances. 
By modelling the experimental data theoretically, and by including diffusion, chemical kinetics, and luminescence, we were able to qualitatively replicate the quasi-linear increase of the CL intensities measured experimentally, as well as to quantitatively reproduce the concentration dependence.

This study demonstrates the potential buried in the free-flowing flat-jet technology for the investigation of liquid-liquid interfaces and interfacial chemical reactions  for identical solvents. 
The flat-jet represents a steady-state system wherein the time-axis is transformed into a spatial coordinate, and imaging provides direct access to time-dependent phenomena.
In contrast to otherwise equivalent experiments using microfluidics, free-flowing jets are not limited by the presence of containers.
These affect the flow dynamics and impose restrictions on the range of wave lengths applicable to spectroscopic studies.
Investigations of liquid–liquid interfaces using free-flowing fluids on the other hand can be extended to X-ray spectroscopies on vacuum flat-jets to access electronic structure, exploiting the unique element specificity and sensitivity to chemical environment of these techniques, and the controlled preparation of interfaces now offers possibilities for the study of, e.g., transfer processes, chemical dynamics, or catalysis.

\section{Acknowledgments}
This work is funded by the SNF (project 200021E-171721) and the DFG in the context of a D-A-CH collaboration, and by the EPFL-MPG doctoral school. We thank Marco Picasso and François Gallaire (both EPFL) for useful discussions. 

\section{Author contribution}
H.C.S., B.W., and A.O. designed the experiments, H.C.S., B.C., A.M.G., and C.O. performed the experiments, and H.H. designed the flat-jet. H.C.S., B.C., A.M.G., and A.O. analyzed the obtained data. H.C.S., B.W., A.M.G., and A.O. wrote the paper, with comments from all authors.

\providecommand{\latin}[1]{#1}
\makeatletter
\providecommand{\doi}
  {\begingroup\let\do\@makeother\dospecials
  \catcode`\{=1 \catcode`\}=2 \doi@aux}
\providecommand{\doi@aux}[1]{\endgroup\texttt{#1}}
\makeatother
\providecommand*\mcitethebibliography{\thebibliography}
\csname @ifundefined\endcsname{endmcitethebibliography}
  {\let\endmcitethebibliography\endthebibliography}{}

\end{document}


\section{A. Luminol reaction pathway resulting in chemiluminescence}
\begin{figure}
\includegraphics[width=0.6\columnwidth]{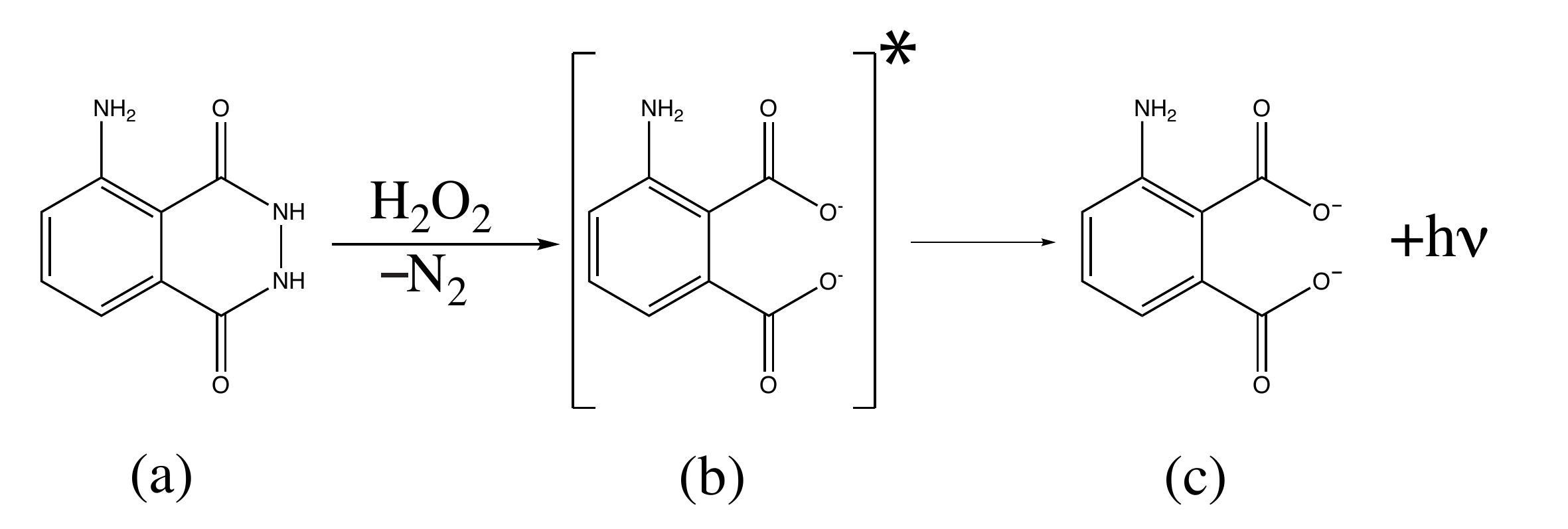} 
\caption{\label{fig:SI1} Reaction of luminol (structure a) with hydrogen peroxide. The latter is decomposed in the presence of a catalyst and forms O$_2$ which then reacts with luminol to form, after loss of N$_2$ structrue b, 3-aminophthalate. The product is formed in an electronically excited state which decays by emission of a photon.
}
\end{figure}
The overall reaction is summarized in figure \ref{fig:SI1}.
5-amino-2,3-dihydro-1,4-phthalazinedione, structure (a) in Figure \ref{fig:SI1} known as Luminol, is oxidized by hydrogen peroxide (\ce{H2O2}) in the presence of a transition metal ion commonly thought to act as a catalyst, though recent work is questioning the veracity of this assumption.\cite{rose1,rose2}
The reaction product, structure (b), is an electronically excited state of the dianion of 3-aminophthalic acid ($AP^*$) which relaxes to the electronic ground state by emitting light at 425 nm. \cite{merenyiEquilibriumReactionLuminol1984,merenyiREACTIVITYSUPEROXIDEO21985,merenyiOxidationPotentialLuminol1990}

This chemiluminescence reaction has been studied in a wide variety of experiments, using metal ions like Cu$^{2+}$, Fe$^{2+}$, Ni$^{2+}$, Co$^{2+}$, hereafter referred to as M$^{2+}$.\cite{rose1} 
Despite all these studies, the reaction mechanism is still not completely understood, and multiple proposals have been explored.\cite{burdo1,gaikwad1,ojima1}
But by now there is broad consent for the mechanism proposed by Merenyi and coworkers.\cite{merenyiEquilibriumReactionLuminol1984,merenyiOxidationPotentialLuminol1990,merenyiREACTIVITYSUPEROXIDEO21985}

The following equations summarize this reaction pathway.
Since \ce{H2O2} is not a sufficiently strong oxidant for luminol,\cite{burdo1,klopf1,rose2} oxidizing radicals must form as a prerequisite, and these are produced through reaction of the metal ion with \ce{H2O2}:
\begin{eqnarray}
\ce{M^{2+} + H2O2  &->&  M^{3+} + OH + OH-} \label{S1}\\
\ce{M^{3+} +H2O2 & -> &   M^{2+} + OH + H+} \label{S1a}\\
\ce{M^{2+} + O2    & -> &  M^{3+} + O_2-} \label{S2}
\end{eqnarray}
In the flat-jet setup, reaction \ref{S1} can only occur when \ce{H2O2} and M$^{2+}$ come into contact via diffusion across the liquid-liquid interface, while reaction \ref{S2} reaction does not require that. 
As such, all three reactions act as part of a parallel recycling pathway where oxidation and reduction of the M$^{2+}$ and M$^{3+}$ ions generate the reactive oxygen containing species.\cite{rose2} 
This suggests that the presence of oxidizing radicals will not be a limiting factor in the reaction network.

The Luminol solution is prepared at a pH$\sim$12, which produces singly deprotonated Lumionol LH$^-$.\cite{augusto1}
In a series of radical reactions
\begin{eqnarray}
\ce{ LH- + OH          & ->&   LH^{\bullet} + OH-} \label{S3}\\
\ce{ LH^{\bullet}  + O2-  & -> &   LOOH-} \label{S4}\\
\ce{ LOOH-                    & -> &   {AP}^{*} + N2(g)\uparrow} \label{S5}\\
\ce{ {AP}^{*}                      & -> &   {AP} + h\nu} \label{S6}
\end{eqnarray}
alpha-hydroxy-hydroperoxide (LOOH$^-$ also known as $\alpha$-HHP), which has been identified as the key intermediate for this reaction, is formed by oxidation of LH$^-$ by different oxygen-containing radicals.\cite{merenyiOxidationPotentialLuminol1990}
Reaction \ref{S5} is the decomposition of $\alpha$-HHP which forms 3-aminophthalic acid in an electronically excited state ($AP^*$) and gaseous \ce{N2}. 
$AP^*$ decays to its ground state by emission of a photon. 

\section{B. Experimental details and sample preparation}
Two high-performance liquid chromatography (HPLC) pumps (Shimadzu, LC-20AD, equipped with degassing unit DGU-20A5R) with matching flow rates around 2.8 ml/min are used to force the solutions through 50-$\mu$m glass capillaries. 
The jets contain, respectively, a 10\% \ce{H2O2} aqueous solution (Merck) and a mixture of Luminol and copper ions in water. 
The second solution is prepared by dissolving 1.2 g \ce{NaHCO3} (Merck, >99.7\%), 0.8 g Luminol (TCI Deutschland GmbH, >98\%), 0.8 g \ce{CuSO4.5H2O} (Roth; purity >99.5\%), and 2.7 g \ce{NH4HCO3} (Merck; purity >99.5\%) in 100 ml water. 
The pH is adjusted to $\sim$12 by addition of NaOH (Merck, >98\%). 
This stock solution is then diluted with water to obtain sample solutions with Luminol concentrations of 6 g/L, 4 g/L, 2 g/L, and 1 g/L, respectively. 

A piezo-operated translation stage was used for the relative positioning of the individual jets, while the collision angle between the jets was fixed at $48^\circ$. 
The resulting first leaf had a length of $\sim$1.5 mm and width of $\sim$0.3 mm, the average flow velocity for both jets was $\sim$22 m/s. 
No measurement of the leaf thickness was done. 

The experiment was conducted under regular atmospheric pressure at room-temperature. 
CL was recorded using a charge-coupled device (CCD) camera that collected the emitted light from the entire leaf. 
Short exposure times were necessary to avoid saturation of the CCD camera when imaging regions with high CL intensity. 
For the results shown in here, 200 images with an individual exposure time of 500 ms were combined. 
Signal background from stray light and CCD dark counts was subtracted by recording identical images under the same conditions but with the jet switched off. 
Images either side of the flat jet yields the same results. 

\section{C. Fluorescence rate}
The lifetime of the excited product state $AP^{*}$ is unknown according to literature, but it is known that $AP^{*}$ decays from of a singlet excited state to the singlet ground state. \cite{augusto1,yue1}
Singlet-singlet transitions in the liquid phase typically take place on time scales faster than $\sim$100 ns,\cite{sauer1} indicating that in our overall rate assessment the fluorescence rate can be neglected.

Further to this general estimation there is strong experimental evidence that the luminescence rate is higher than the temporal resolution of the experiment. 
Figure 3B of the main manuscript shows the CL intensity as a function of the downstream coordinate $y$. 
At $y\sim -200\pm200$ pixels, there is an intense CL peak that decays to zero. 
We attribute the CL spike to turbulent mixing at the point where the two jets collide.

This observation suggests that the luminescence is emitted precisely where the solutions mix. 
If, on the other hand, the luminescence would be slower then the bright peak would be smeared over a larger area downstream from the point where the jets first meet.
Thus, if there would be predominately phosphorescence from a long-lived triplet state, the experiment would have shown nearly constant luminescence throughout the entire first leaf. 

The time resolution of the experiment is estimated as follows: 
During the total exposure time of 100 seconds (see above), the jet and camera slightly vibrate.
These vibrations have an estimated amplitude of estimate to be 30 -- 50 $\mu$m, resulting in a blurring over $\sim$27 -- 45 pixels. 
The downstream velocity of 23 m/s implies that the temporal resolution is around 2.16 $\mu$s. 

\begin{figure}[H]
    \centering
    \includegraphics[width=75mm]{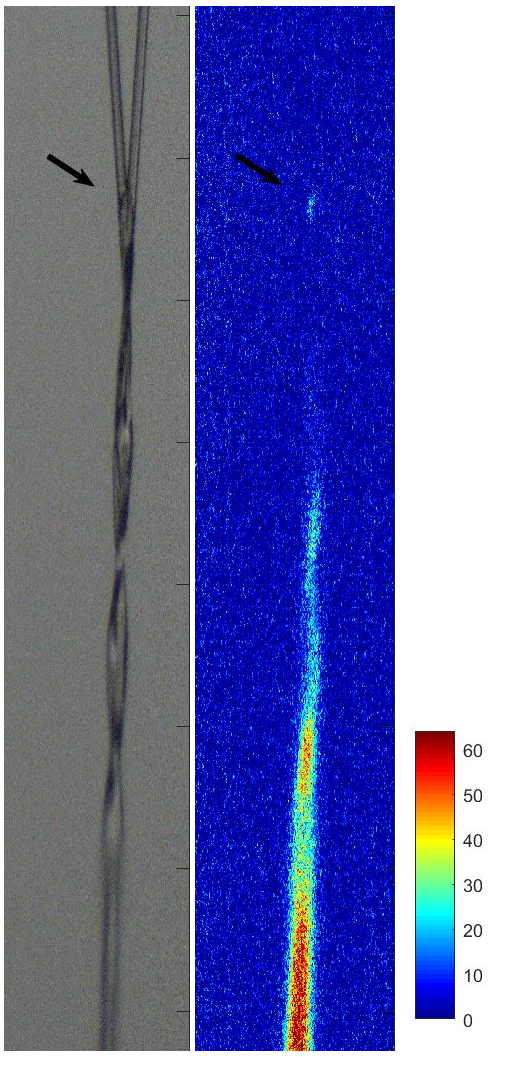}
    \caption{Left part: CCD image of the two intentionally misaligned single jets resulting in a spiral. Right part: false color image of the luminescence intensity. Black arrow indicates the position where the two jets touch and start to self-spiral.}
    \label{fig:badjet}
\end{figure}
We further investigated the phenomenon around the impinging point of the two single jets. 
The left panel of figure \ref{fig:badjet} depicts a CCD image where the two single jets have intentionally been misaligned relative to each other in such a way as to form a helical flow structure that starts with a point of impact at the top of the photo, as indicated by a black arrow.
The right panel of Figure \ref{fig:badjet} shows a false color plot of the luminescence intensity resulting from this arrangement.
We observe a small luminescence peak at the point of first contact (black arrow) which we attribute to turbulent mixing where the two single jets touch. 
The luminescence intensity dies out completely and only slowly increases further downstream. 
This confirms that the luminescence is emitted exactly where mixing takes place; the lifetime of the excited state seems to be short enough to not lead to a delayed luminescence. 
Indeed, a long-lived triplet state would lead to a constant luminescence intensity all the way downstream along the structure, rather than the localised peak. 

It is important to note the difference between the flat jet leaf structure and the present arrangement. 
In the former case the rims of the leaves exhibit turbulent mixing and thus luminesce along the entire first leaf. 
With that experiment alone one can not exclude that the luminescence observed in those regions results from reactions taking place at the point of impact.
In contrast, the present structure clearly shows complete disappearance of the luminescence after the first impact which allows us to conclude that the luminescence observed in the rims of the leaves results from local mixing in those domains, rather than from luminescence from prior reactions.

\section{D. Diffusivity and concentration gradients}
The mechanism bringing the reactants from the two solutions into contact is diffusion across the liquid-liquid interface.
Diffusion for the relevant species is characterised by the diffusion coefficient, as listed in table \ref{tab:diffconst}.

\begin{table}[]
    \centering
    \begin{tabular}{|c|c|c|}
    \hline
    Reactant species & Diffusion coefficient & Reference \\
    & [m$^2$/s] &\\  \hline
    H$_2$O$_2$   &  17.5$\cdot$10$^{-10}$ &\citep{gros1}\\
    Luminol       &  13.5$\cdot$10$^{-10}$ &\citep{koizumi1} \\ 
    Cu$^{2+}$    &   7.1$\cdot$10$^{-10}$& \citep{vanysek1}\\
    O$_2$       &  20.0$\cdot$10$^{-10}$  & \citep{xing1}\\
    OH$^-$       &  52.7$\cdot$10$^{-10}$ & \citep{vanysek1}\\ \hline
         
    \end{tabular}
    \caption{diffusion coefficients for the reacting species}
    \label{tab:diffconst}
\end{table}

The concentration distribution c($x,t$) is calculated using Fick's law.\cite{steinfeld1}
All reactant species are dissolved at rather low concentrations in aqueous solution and the flow-dynamics are laminar, justifying this choice. 
The two solution flat-jet is modelled using the concentration distributions at the origin c($x$=0,$t$=0) as a step-function with uniform non-zero concentration on one side of the interface, zero concentration on the other side, and the interface to be infinitely thin. 
Such circumstances can be modelled well using the error-function\footnote[1]{Note that because of symmetry reasons we are using the error-function defined as$^{18}$ ${\displaystyle \operatorname {erf} z = \Gamma(z) ={\frac {2}{\sqrt {\pi }}}\int _{0}^{z}e^{-t^{2}}\,dt}$ and not the sometimes-used definition of a cumulative distribution function of the normal distribution function. The two definitions are equivalent except for a linear vertical shift, which would require a different coefficient for this model to have physical meaning.}, hereafter denoted with $\Gamma$:
\begin{eqnarray}
    c_{i}(x,t) & = & \frac{c_{i,0}}{2}\left[1\pm\Gamma \left(\frac{x}{2 \sqrt{D_i\:t}} \right) \right] \label{S10}
\end{eqnarray}
where $c_{i,0}$ and $D_{i}$ are the concentration and diffusion coefficient of species $i$, respectively, and $t$ defines the temporal coordinate. 
The last term is negative for one solution and positive for the other, indicating that initially the concentration on the other side of the interface is zero.
The expressions `1-$\Gamma$(...)' and `1+$\Gamma$(...)' ensure that the concentration distribution only takes values between zero and  $c_0$. 
Note that the temporal and spatial origin $t$ = $x$ = $y$ = $z$ = 0 does not correspond to the moment and the position where the two cylindrical jets collide, but to the point (and time) where the luminescence has returned to zero after the initial spike, as is indicated in Figure 2 of the main manuscript. 

\begin{figure}[H]
    \centering
    \includegraphics[width=0.8\linewidth]{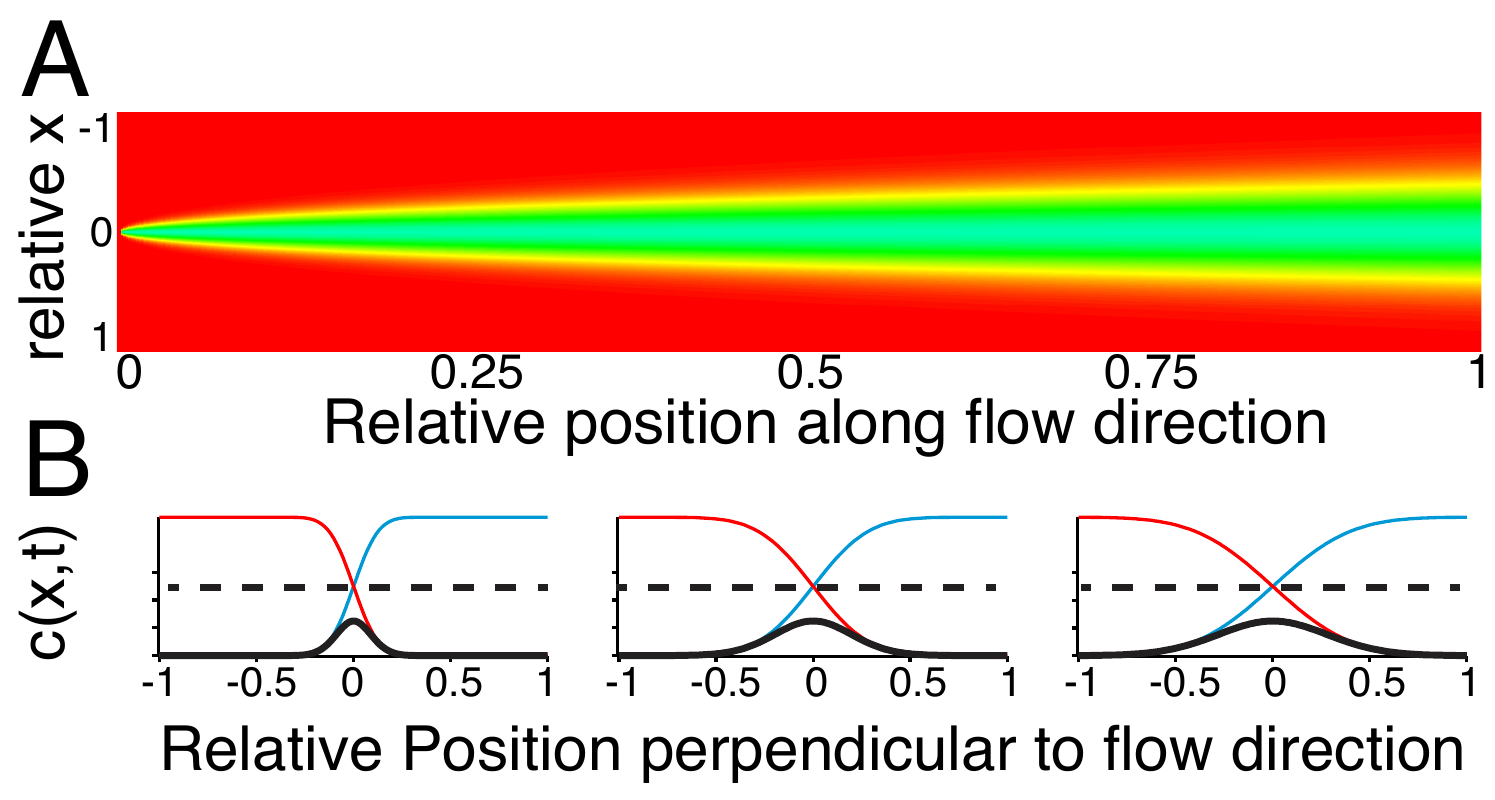}
    \caption{A) Overlap of the product concentrations $c_i(x,t)$ for two species with equal diffusion coefficients and equal initual concentration to illustrate the evolution of the reactive layer in the flat-jet. The concentration profiles shown in a cross section through the flat-jet leaf along the propagation axis. B) Concentration distributions $c(x,t)$ of the species from the left (red) and right (blue) sheet of the leaf as a function of time, which is equivalent to downstream position. The graphs show the distributions, from left to right, at relative positions 0.25, 0.5, and 1, respectively. Equal diffusion coefficients are assumed for both species. The black line is the product of the red and blue traces and hence is proportional to the reaction rate of a reaction with first order kinetics in the reactants contained in either solution. The horizontal scale is the distance perpendicular to the jet propagation and the leaf surface, $x$.}
    \label{fig:SI-diffusion}
\end{figure}

Figure \ref{fig:SI-diffusion}B shows the evolution of the concentration profiles with time.
Since in our experiment time is equivalent to the position downstream of the jet, these profiles correspond to different positions along the $y$ coordinate.
Profiles are given along $x$, i.e. perpendicular to $y$ and the leaf surface.
We here only give arbitrary units since the relation between $t$ and $y$ is the flow velocity which is an experimental parameter, and the absolute widths of the concentration profiles depend on the specific species and solvents.
In the present experiment, the width of the overlapping layer after a flow distance of 1 mm corresponds to $\sim$200 nm.

\section{E. Model of the diffusion-limited reaction and CL}
Following reaction equations \ref{S1} to \ref{S4}, the rate of formation of $AP^*$ is written as a function of the limiting reagents:
\begin{eqnarray}
    \frac{d[\ce{AP}^*]}{dt} & = & k_{dec}[\ce{LOOH}] \\ \label{S7}
    & = & k_{dec}\:k_{rad}[\ce{H_2O_2}][\ce{LH^-}]  \label{S8}
\end{eqnarray}
where k$_{dec}$ and k$_{rad}$ respectively describe the rate coefficients for the decomposition and the various steps within the radical reaction. 
Note that the Luminol concentration is writen as $[LH^-]$ because high pH of the Luminol solution implies that is only present only in the singly deprotonated form. 
Since in our experiment we only observe the decay of the final product, we contract the these steps into one single reaction with rate coefficient k$_{R}$ = k$_{dec}$ k$_{rad}$ and rewrite equation \ref{S8} accordingly
\begin{equation}
    \frac{d[\ce{AP}^*]}{dt} = k_R[\ce{H_2O_2}][\ce{LH^-}]   \label{S9}
\end{equation}

The two relevant spatial dimensions are along the jet axis and perpendicular to the leaf surface, which are denoted by $y$ and $x$, respectively. 
Figure \ref{fig:SI-diffusion}A shows the overlap of the concentration profiles as a function of relative downstream position (horizontal axis) and perpendicular to the leaf surface (vertical axis).
This plot is critical for the understanding of the relevant dynamics, since it shows the product of the two concentration profiles which, for a kinetics that is of first order in either reactant, is proportional to the reaction rate.
Equation \ref{S9} can thus be rewritten as
\begin{eqnarray}
    \frac{d[\ce{AP}^*]}{dt}  =  k_R \Biggl\{\frac{[\ce{H2O2}]_0}{2}\left[1-\Gamma\left(\frac{x}{2\sqrt{D_{\ce{H2O2}}t}}\right)\right] \times \frac{[\ce{LH^{-}}]_0}{2}\left[1+\Gamma\left(\frac{x}{2\sqrt{D_{\ce{LH^{-}}}t}}\right)\right] \Biggr\}  \label{S11}
\end{eqnarray}
which explicitly takes into account the concentration profiles for the two diffusing reactants. 
This rate is valid for any position along the cross section of the jet, at any given time and thus position along the flat part of the first leaf. 
The total product concentration -- here temporarily assuming the product stays in its excited state and does not decay -- as a function of downstream time is
\begin{eqnarray}
    [\ce{AP}^*] & = & \int_0^{\tau} k_R\int_{-\delta_L/2}^{\delta_L/2} \Biggl\{\frac{[\ce{H2O2}]_0}{2}\left[1-\Gamma\left(\frac{x}{2\sqrt{D_{\ce{H2O2}}t}}\right)\right] \nonumber \times\\
    &  & \frac{[\ce{LH^{-}}]_0}{2}\left[1+\Gamma\left(\frac{x}{2\sqrt{D_{\ce{LH^{-}}}t}}\right)\right] \Biggr\}dx\: dt   \label{S12}
\end{eqnarray}
or, as function of the downstream coordinate $y$, using $t$ = $y/v$:
\begin{eqnarray}
    [\ce{AP}^*] & = & \int_0^{\psi} \frac{k_R}{v}\int_{-\delta_L/2}^{\delta_L/2} \Biggl\{\frac{[\ce{H2O2}]_0}{2}\left[1-\Gamma\left(\frac{x}{2\sqrt{D_{\ce{H2O2}}y/v}}\right)\right] \nonumber \times\\
    & & \frac{[\ce{HL^{-}}]_0}{2}\left[1+\Gamma\left(\frac{x}{2\sqrt{D_{\ce{HL^{-}}}y/v}}\right)\right] \Biggr\}dx\: dy   \label{S13}
\end{eqnarray}
The integrals correspond to integration along the vertical and the horizontal axis in the graph of figure \ref{fig:SI-diffusion}A.

Following this, the fluorescence at time $t$ = $\tau$ is calculated from the integrated build-up of [$AP^*$] along the jet propagation minus the part that decays through CL. However, it is favorable to use a differential form, reflecting the rate of increase of concentration of $AP^*$ through the reaction and the rate of reduction through CL and again written in terms of the downstream position $y$:
\begin{eqnarray}
    \frac{\Delta[\ce{AP}^*]}{\Delta y} & = & \frac{1}{v}\Biggl[-k_P[\ce{AP}^*]\nonumber + k_R\int_{-\delta_L/2}^{\delta_L/2} \Biggl\{\frac{[\ce{H2O2}]_0}{2}\left[1-\Gamma\left(\frac{x}{2\sqrt{D_{\ce{H2O2}}y/v}}\right)\right] \nonumber \times\\
    & &  \frac{[\ce{HL^{-}}]_0}{2}\left[1+\Gamma\left(\frac{x}{2\sqrt{D_{\ce{HL^{-}}}y/v}}\right)\right]\Biggr\}dx\Biggr]   \label{S15}
\end{eqnarray}
For simplicity the overlap integral is abbreviated as $\shin$: 
\begin{eqnarray}
    \shin(y) & = & \int_{-\delta_L/2}^{\delta_L/2} \Biggl\{\frac{[\ce{H2O2}]_0}{2}\left[1-\Gamma\left(\frac{x}{2\sqrt{D_{\ce{H2O2}}y/v}}\right)\right] \nonumber \times\\
    &  & \frac{[\ce{HL^{-}}]_0}{2}\left[1+\Gamma\left(\frac{x}{2\sqrt{D_{\ce{HL^{-}}}y/v}}\right)\right]\Biggr\}dx    \label{S16}
\end{eqnarray}
leading to equation 4 of the main manuscript.

There is no closed-form solution to this differential equation since it contains the integral of the product of two error functions, and the differential equation was solved numerically. 
The rate of photon emission is given by $\Phi = k_P [AP^*](t)$ (equation 1 of the main manuscript). 
Thus, the number of photons per unit time was calculated from the time- and position-dependent formation of $AP^*$. 
The entire dynamics given by equation \ref{S15} were simulated by stepwise numerical integration of the CL rate using both Euler’s method and RK4.\cite{cheney1} 
The two algorithms give rise to the same good agreement, indicating that the calculations have converged. 

Three possible solutions resulting from the numerical integration of this expression under different initial conditions are presented in in the inset of figure 3B in the main paper. 
These three exemplary cases highlight possible behaviors for the CL profiles, with the different initial conditions arising due to the fact that the exact values of the kinetic coefficients k$_P$ and k$_R$ (and even the orders of magnitude) are unknown for this system. 
In addition, we do not have an absolute count of number of photons due to the fact that the imaged solid angle is unknown as well as the quantum efficiency from the CCD chip.

\section{F. Fitting procedure}
In the following we describe how the measured CL intensities as a function of the spatial downstream coordinate \textit{y} can be fitted using a combination of numerical and regression techniques. Typically one would use as input for a regression routine $\boldsymbol{X}_{data}$, $\boldsymbol{Y}_{data}$ and a selected function \textit{F} that relates $\boldsymbol{Y}_{data}$ = $F(\boldsymbol{X}_{data}, \boldsymbol{p})$ where \textit{F} contains some parameters $\boldsymbol{p}$ that the routine optimizes to achieve a best fit according to an optimization algorithm for the input data-set. The equation to fit to is the differential form given in the main text by equation 4, where we have a general formula for $d[AP^*](y)/dy$ as a function of $[AP^*](y)$ and the concentration gradient overlap $\shin(y)$, equation \ref{S16}. Despite having neither the absolute excited state concentration nor its spatial derivative, we do have the detector signal V(\textit{y}) as a function of the downstream coordinate \textit{y}, which we can use as a proxy for the photon flux $\Phi(y)$ using a linear transform of the integrated photon count. We use a linear transform to describe the relation between V(\textit{y}) and $\Phi(y)$ to allocate additional flexibility in the fit and to correct for calibration in generalized future experiments:
\begin{eqnarray}
    \text{V}(y) & = & \alpha \: \Phi(y) + \beta \\ \label{S18a}
    & = & \alpha \: k_P [AP^*](y) + \beta \label{S18b}
\end{eqnarray}
where in the second step, equation 1 of the main manuscript: $\Phi(y) = k_P[AP^*](y)]$ is applied. The parameters $\alpha$ and $\beta$ represent the linear attenuation factor and an offset respectively, which account first for the fraction of the photon flux $\Phi(y)$, and secondly for the quantum efficiency of the CCD chip itself. Solving this equation for $[AP^*](y)$ gives:
\begin{equation}
   [AP^*](y) = \frac{1}{\alpha \: k_P} (\text{V}(y) - \beta) \label{S19a}
\end{equation}
and its differential form:
\begin{equation}
    \frac{\Delta[AP^*](y)}{\Delta y} = \frac{1}{\alpha \: k_P} \frac{\Delta \text{V}(y)}{\Delta y} \label{S19b}
\end{equation}

Substituting equations \ref{S19a} and \ref{S19b} into equation 4 of the main text and simplifying allows us to write our differential equation model completely in terms of the detected intensity instead of the excited state product concentration. This transformation takes the molecular-scale ODE and allows it to be applied to our observable, the detector intensity:
\begin{equation}
    \frac{\Delta \text{V}(y)}{\Delta y} = \frac{1}{v}\left[-k_P \: \text{V}(y) + \alpha\:k_P\:k_R\: \shin(y) + k_P \: \beta \right]
\end{equation}

\begin{figure}[]
    \centering
    \includegraphics[width=130mm]{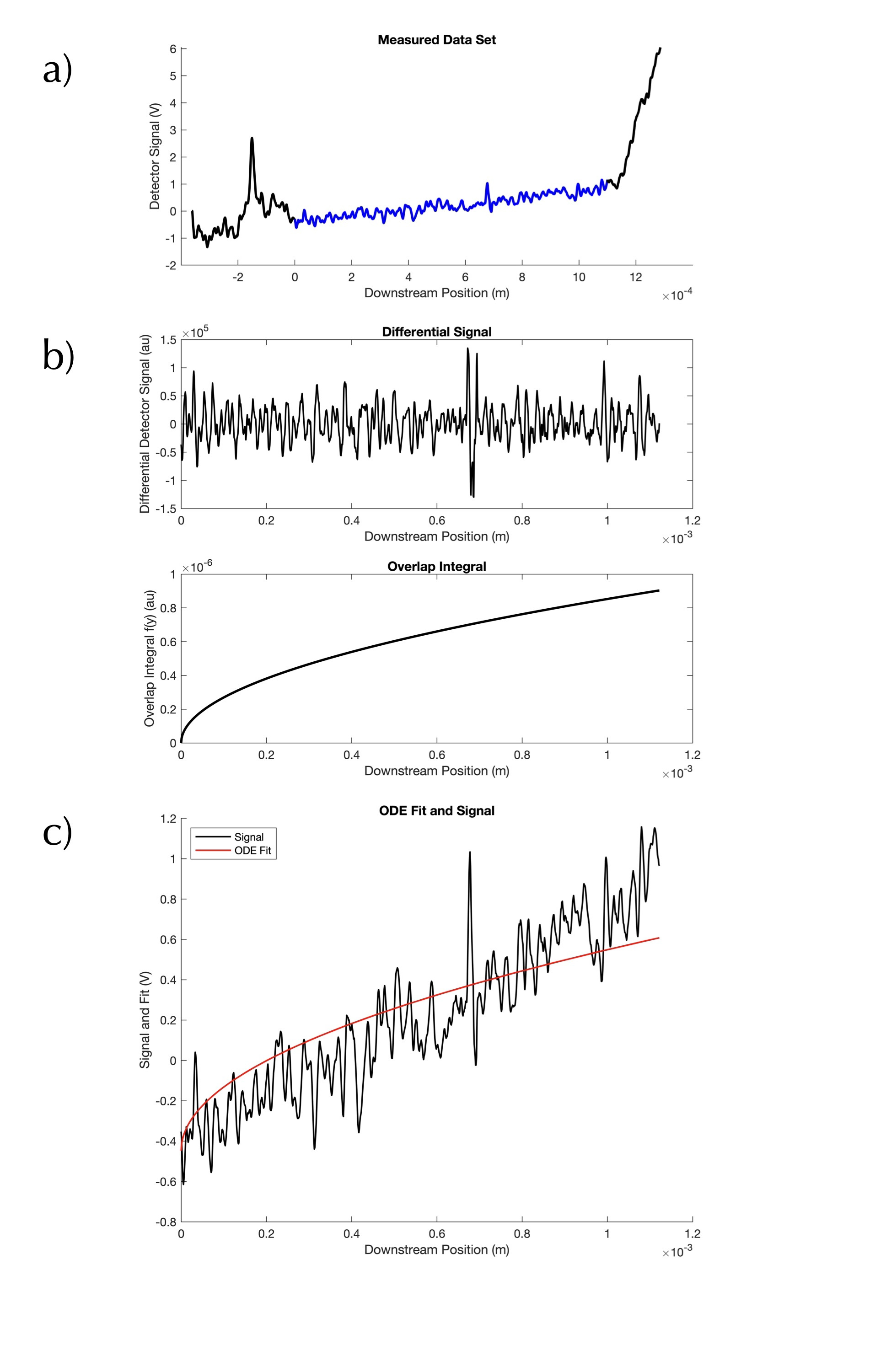}
    \caption{Overview of the fitting procedure: a) shows the measured data of the 1g/L concentrated solution in black and the data interval selected for the fitting in blue. b) the basis vectors used in the regression model: the differential signal (top) and numerically-computed overlap integral (bottom). c) The output from the ODE regression model is plotted in red against the signal shown in part a).}
    \label{fig:SI-3}
\end{figure}

Fitting with differential data as the response variable leads to an inherent loss of information due to the missing constant terms, so it is advantageous to solve this ODE for the detected intensity in terms of the differential:
\begin{eqnarray}
    \frac{k_P}{v} \text{V}(y) & = &  -\frac{\Delta \text{V}(y)}{\Delta y} + \frac{\alpha\:k_P\:k_R}{v}\shin(y) + \frac{k_P \: \beta}{v} \\
    \text{V}(y) & = & -v \frac{1}{k_P} \frac{\Delta \text{V}(y)}{\Delta y} + \alpha\:k_R\:\shin(y) + \beta
\end{eqnarray}
We then introduce dummy variables to allow for the regression model to converge without needing to resolve the nonlinearly correlated parameters $\alpha$ and $k_R$:
\begin{equation}
    \text{V}(y) = -v\:b_1 \frac{\Delta \text{V}(y)}{\Delta y} + b_2\:\shin(y) + b_3\label{eq:26}
\end{equation}
where $b_1 = 1/k_P$ is the excited-state lifetime, $b_2 = \alpha \:k_R$, and $b_3 = \beta$. After numerically calculating $\shin(y)$ using the diffusion constants listed in table SI-1 we feed $\shin(y)$ and the measured differential data $\Delta \text{V}(y)/\Delta y$ as independent variables and the signal $\text{V}(y)$ as the response variable into the fitting routine using the differential equation as the model, equation \ref{eq:26}. 
Figure \ref{fig:SI-3} illustrates the fitting procedure: part a) shows the measured data of the 1g/L concentrated solution as the black line and the selected subinterval for the fitting (the region for which this model is applicable) as the blue line. Part b) depicts the two linearly-independent basis vectors used in the regression model: top is the numerical differential of the signal and bottom is the numerically-computed overlap integral. These two basis vectors are used to reproduce the signal vector $\text{V}(y)$ shown in part a) by means of least squares minimization. Part c) shows the output of the regression model as the optimal predicted signal given the basis vectors input. It is plotted in red together with the data region highlighted in part a).\\\\

While Figure \ref{fig:SI-3} shows the fit process for a single concentration for the sake of clarity, the actual fitting process uses all four concentrations in a single fit. 
This is accomplished by concatenating each of the generated basis vectors together and produces a statistically superior fit compared to fitting each concentration independently. 
Considering all three fit-parameters and their correspondence to the physical parameters, we will first discuss the radiative decay which follows from $b_1 = 1/k_P$, and will analyze the implications of the other two resultant parameters later. 
The output of a standard regression algorithm under the constraint that $b_1$ is non-negative is that $b_1 = 2.248 \times 10^{-14}$ \textit{s} with a 95\% confidence interval of $[-1.961\times10^{-8}, 1.961\times10^{-8}]$ \textit{s}. 
We can use this method of determining the parameter confidence intervals because the parameters $b_i$ are linearly independent.$^{19}$ 
This corresponds to a $k_P$ of $4.448 \times 10^{13} \:s^{-1}$ with a 95\% confidence interval of $[-5.099\times10^7, 5.099\times10^7] \:s^{-1}$. $b_1 = 2.248 \times 10^{-14}$ \textit{s} is an excited-state lifetime of roughly 20 fs. 

\begin{figure}[]
    \centering
    \includegraphics[width=110mm]{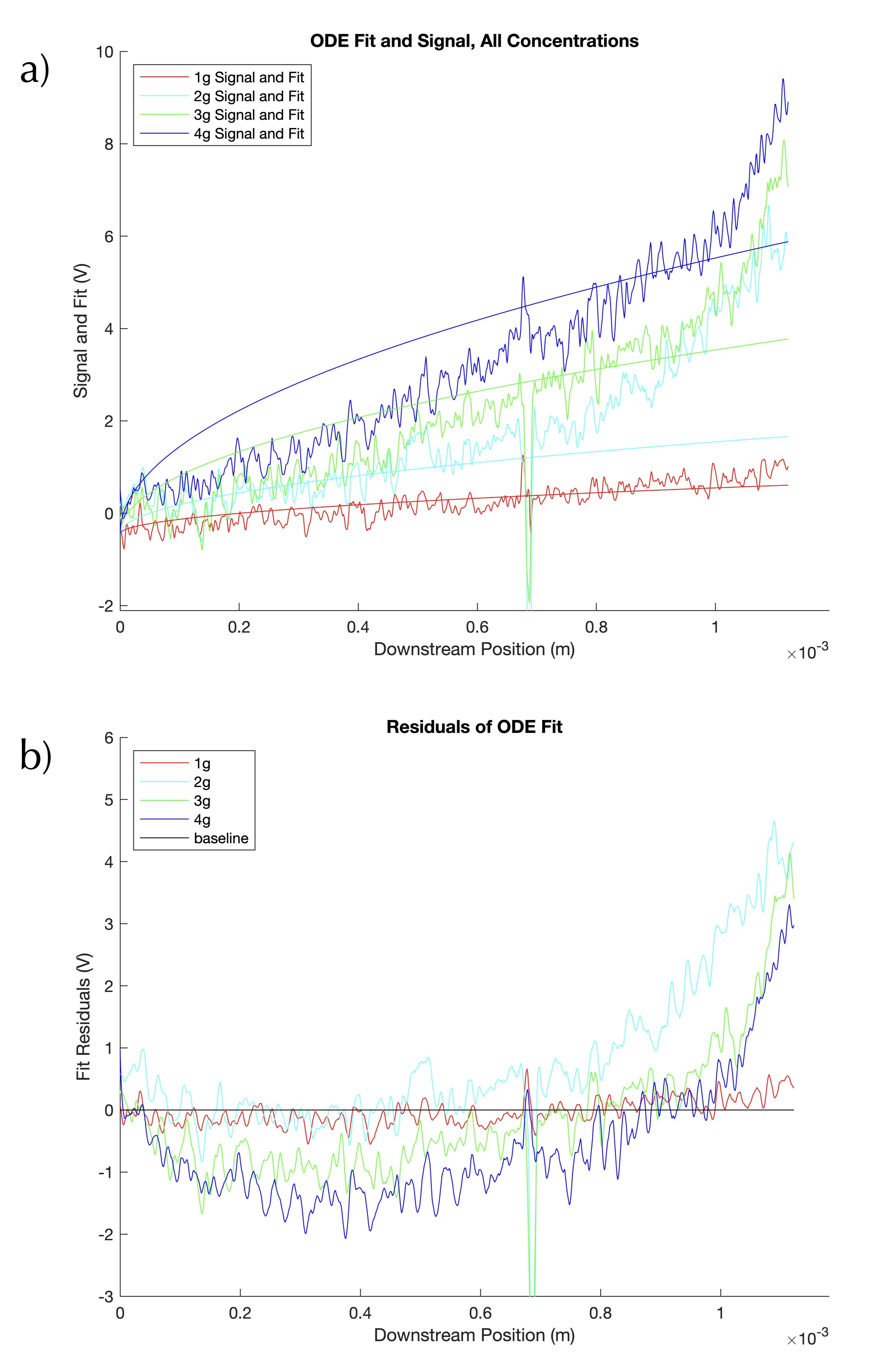}
    \caption{a) Signal and simultaneous ODE regression fits for all four concentrations. Note how the behavior of each of the fits is dominated by the overlap integral plotted in Figure \ref{fig:SI-3}b. b) Residuals of the regression shown in a. Note the clear pattern in the residuals, indicating possible issues with the use of the differential data as a basis vector or the need for a different model.}
    \label{fig:SI-4}
\end{figure}

Evaluating the model with the optimal parameters and all four concentrations yields part a of Figure \ref{fig:SI-4}. 
This shows that the model does replicate the general behavior of the chemiluminescence traces, but fails to properly predict the linear increase. 
Instead, the behavior of the evaluated fits is dominated by the overlap integral, which makes sense given that the differential signal behaves as a constant that is approximately zero; the contribution from this basis vector is unsurprisingly negligible. 
This difference between the optimal predicted chemiluminescence trace and the actual data is best visualized as a residual plot, shown as part b of Figure \ref{fig:SI-4}. An appropriate regression output would have this plot be featureless with no discernible pattern about the V = 0 baseline. 
Instead, we see that the model systemically under-predicts the signal at small downstream positions and over-predicts the signal at large downstream positions. 
This often means that the model is not appropriate for regression and/or does not suit the data well, but we have sufficient reason to attribute this to two other factors in the measured data and experimental setup:

1.) The fit results for $b_1$ and the corresponding lifetime vary around values on the order of 20 ns. 
This value is nearly a factor 1000 less than the temporal resolution provided by our experiment and hence can not be considered physically meaningful.
It does, however, imply that the luminescence rate is sufficiently fast to not affect the evolution of the observed luminescence profiles because the excited state lifetime is so short that the photon is emitted nearly exactly at the position where the product structure is formed.

2.) The values of first spatial derivative is close to zero, which can well be observed in Figure \ref{fig:SI-3}b). This means that only few additional data points within the fitted interval can have a drastic influence on the fit-result. Moreover, we have two parts within the fit-function: $-k_P$ times the differential signal describes the approximately linear decay and $k_R$ times the overlap integral that induces an approximately linear growth (for certain regimes of the overlap integral). As such, there are a number of possible combinations of parameters that could model the signal well and the combination that gives rise to the least-squares minimum may not be what actually corresponds with the molecular parameters that produced that curve.

Obtaining $b_3 = \beta = -0.4469$ merely indicates that the signal we observe has a negative vertical shift from V = 0 which is predicted by the model. 
However, obtaining information about about $k_R$ from the last dummy parameter $b_2$ is considerably more difficult. 
The  parameters $\alpha$ and $k_R$ that define $b_2$ are correlated and there currently is no way to correct for this using the covariance matrix.

Even still, this method stands on its own as a demonstrative example of how quantitative chemical kinetics information can be extracted from a flat-jet system such as the one we show here. We show this method not to reveal any novel scientific information about the luminol system we present in this manuscript but as a technical detail to those who wish to replicate these techniques in other systems. Based on the above discussion, it is reasonable to conclude that this process would indeed be effective in extracting molecular kinetics information, even from traditionally-difficult kinetics systems such as luminol and hydrogen peroxide, so long as the detector has significantly-improved spatial resolution and is calibrated. This method also has the potential too be quite effective on systems with longer excited state lifetimes so that experiemental capabilities can better resolve the chemiluminescence. Even as-is, this analysis - in particular the confidence interval/ hypothesis test for $k_P$ - has statistically shown that quantitative information should not be extracted/interpreted from this data set, itself an extremely useful result.

\bibliographystyle{achemso}
\providecommand{\latin}[1]{#1}
\makeatletter
\providecommand{\doi}
  {\begingroup\let\do\@makeother\dospecials
  \catcode`\{=1 \catcode`\}=2 \doi@aux}
\providecommand{\doi@aux}[1]{\endgroup\texttt{#1}}
\makeatother
\providecommand*\mcitethebibliography{\thebibliography}
\csname @ifundefined\endcsname{endmcitethebibliography}
  {\let\endmcitethebibliography\endthebibliography}{}